\documentclass[reprint,aps,superscriptaddress]{revtex4-2}
\pdfoutput=1
\usepackage{graphicx}
\usepackage{tcolorbox}
\usepackage{bm}
\usepackage{xr}
\usepackage{float}
\usepackage{amsmath}
\usepackage{booktabs}
\usepackage{xcolor}
\usepackage{soul}
\usepackage{cleveref}
\usepackage{circledsteps}
\usepackage[normalem]{ulem}

\begin{document}

\title{Higher-order motif analysis in hypergraphs}

\author{Quintino Francesco Lotito}
\affiliation{Department of Information Engineering and Computer Science, University of Trento, via Sommarive 9, 38123 Trento, Italy}

\author{Federico Musciotto}
\affiliation{Dipartimento di Fisica e Chimica Emilio Segr\`e, Universit\`a di Palermo, Viale delle Scienze, Ed. 18, I-90128, Palermo, Italy}

\author{Alberto Montresor}
\affiliation{Department of Information Engineering and Computer Science, University of Trento, via Sommarive 9, 38123 Trento, Italy}

\author{Federico Battiston}
\affiliation{Department of Network and Data Science, Central European University, 1100 Vienna, Austria}

\begin{abstract}
A deluge of new data on social, technological and biological networked systems suggests that a large number of interactions among system units are not limited to pairs, but rather involve a higher number of nodes. To properly encode such higher-order interactions, richer mathematical frameworks such as hypergraphs are needed, where hyperlinks describe connections among an arbitrary number of nodes. Here we introduce the concept of higher-order motifs, small connected subgraphs where vertices may be linked by interactions of any order. We provide lower and upper bounds on the number of higher-order motifs as a function of the motif size, and propose an efficient algorithm to extract complete higher-order motif profiles from empirical data. We identify different families of hypergraphs, characterized by distinct higher-order connectivity patterns at the local scale. We also capture evidences of structural reinforcement, a mechanism that associates higher strengths of higher-order interactions for the nodes that interact more at the pairwise level. Our work highlights the informative power of higher-order motifs, providing a first way to extract higher-order fingerprints in hypergraphs at the network microscale.
\end{abstract}

\maketitle

\section*{Introduction}

Over the last two decades, \emph{networks} have emerged as a powerful tool to analyse the complex topology of interacting systems~\cite{boccaletti2006complex}.
From social networks to the brain, several systems have been represented as a collection of nodes and links, encoding dyadic connections among pairs of units. Yet, growing empirical evidence is now suggesting that a large number of such interactions are not limited to pairs, but rather occur in larger groups~\cite{battiston2020networks}. Examples include collaboration networks~\cite{patania2017shape}, human face-to-face interactions~\cite{cencetti2021temporal}, species interactions in complex ecosystems~\cite{grilli2017higher} and structural and functional brain networks~\cite{petri2014homological,giusti2016two}. 

To properly encode such \emph{higher-order interactions}~\cite{battiston2020networks}, richer mathematical frameworks are needed such as hypergraphs~\cite{berge1973graphs}, where hyperedges describe interactions taking place among an arbitrary number of nodes. To characterize these \emph{higher-order systems}~\cite{battiston2020networks}, computational tools from algebraic topology have been proposed~\cite{patania2017topological,sizemore2019importance}, as well as generalization of common network concepts, including centrality measures~\cite{estrada2006subgraph, benson2019three}, clustering~\cite{benson2018simplicial, yin2018higher} and assortativity~\cite{veldt2021higher}. An explicit treatment of higher-order interactions, including their inference and reconstruction~\cite{young2020hypergraph}, is necessary to understand network formation mechanisms~\cite{courtney2016generalized,chodrow2020configuration,kovalenko2021growing,millan2021local}, fully capture the real community structure of higher-order systems~\cite{carletti2021random, eriksson2021choosing, chodrow2021generative} and extract their statistically validated higher-order backbone~\cite{musciotto2021detecting}. Noticeably, taking into account higher-order interactions might be crucial to understand the emergent behavior of complex systems, as they have been found to profoundly impact
diffusion~\cite{schaub2020random,carletti2020random}, synchronization~\cite{bick2016chaos, skardal2020higher, millan2020explosive, lucas2020multiorder,gambuzza2021stability}, social~\cite{iacopini2019simplicial, chowdhary2021simplicial, neuhauser2020opinion}  and evolutionary~\cite{alvarez2021evolutionary} processes.

Networked systems may be differentiated by their preferential patterns of connectivity at the microscale, encoding a characteristic fingerprint often relevant for the system function. This may be quantified by measuring network \emph{motifs}, small connected subgraphs that appear in an observed network at a frequency that is significantly higher than in its randomized counterpart~\cite{milo2002network}. The analysis of the motifs of a network revealed the emergence of ``superfamilies'' of networks, i.e. clusters of networks which display similar local structure. These clusters tend to group networks from similar domains or networks that undertook similar evolutionary processes~\cite{milo2004superfamilies}. 

Motifs can be interpreted as elementary computational circuits, with specific functionalities that can be shared by similar networks. For example, transportation networks are designed to simplify the traffic flow, whereas gene regulation and neuron networks are evolved to process information. In this regard, studying motifs can also give new insights on the dynamics and resilience of classes of networks~\cite{milo2004superfamilies, Dey2019resilience}. To explicitly uncover the relation between the dynamical processes that unfold on a network and its structural decomposition at the local scale, recently a refined notion of \emph{process motifs} has been proposed~\cite{schwarze2020motifs}, introducing a framework to assess the contribution of each motif to the overall dynamical behavior of the system.  

Network motifs have been used in a wide range of applications. In biology, motifs have been extensively studied for the analysis of transcription regulation networks (i.e. networks that control gene expression). Findings show that diverse organisms from bacteria to humans exhibit common regulation patterns, each with its very own function in determining gene expression~\cite{alon2007network, shen2002network, mazurie2005evolutionary, dobrin2004aggregation, yeger2004network}. Similarly, motif analysis has been applied to show how complex and flexible neural functions emerge from the composition of fundamental circuits in brain networks ~\cite{sporns2004motifs}. Moreover, motifs have also been used as a feature for the identification of cancer~\cite{chen2013identification}. Eventually, the need for biological applications to analyze datasets of ever-increasing size has been a strong motivation for the development of more efficient algorithms~\cite{Sabyasachi2020review}. Beside biology, motifs have also been applied to provide fingerprints of the local structures of social networks~\cite{hong2014social, juszczyszyn2008local}, for the early detection of crisis-leading structural changes in financial networks~\cite{saracco2016detecting} and to study the networks of direct and indirect interactions across species in ecology~\cite{bascompte2009assembly, simmons2019motifs}.

The interest of the research community in extracting fingerprints at the network microscale of real-world systems has led to considering richer frameworks for motif analysis~\cite{benson2016higher}, including extensions to more general network models such as weighted~\cite{barrat2004networked}, temporal~\cite{holme2012temporal} and multilayer~\cite{battiston2014structural} networks. Weighted networks can be characterized in terms of the intensity and coherence of the link weights of their subgraphs~\cite{onnela2005intensity}. Temporal networks can be studied at both topological and temporal micro and mesoscale by considering time-restricted patterns of interactions~\cite{kovanen2011temporal, paranjape2017motifs}. Finally, statistically over-represented small multilayer subgraphs~\cite{kivela2018isomorphism} highlight the local structure of multilayer networks such as the human brain~\cite{battiston2017multilayer}.  

%- what is missing in the literature.

%- OUR CONTRIBUTION: higher-order motifs for hypergraphs, What we do / find / outline of the paper

The methods, algorithms and tools proposed in literature so far only consider patterns of pairwise interactions, thus limiting our capabilities of characterizing the local structure of systems that involve group interactions. In order to fill this gap, we introduce the notion of higher-order network motifs in hypergraphs, which are defined as statistically over-represented connected subgraphs with higher-order interactions. We propose a combinatorial characterization of these new mathematical objects and develop an efficient algorithm to evaluate the statistical-significance of each higher-order motif on empirical data. 
We show that we are able to extract fingerprints at the network microscale of higher-order real-world systems, and highlight the emerge of families of systems that show a similar higher-order local structure. Finally, we observe the phenomenon of structural reinforcement, for which real-world group interactions take place more frequently if they are supported by a rich hierarchical structure of pairwise interactions.

\section*{Results}

Motif analysis is established as a fundamental tool in network science to extract fingerprints of networks at the microscale and to identify their structural and functional building blocks. By directly extending the traditional definition of network motifs, we can define higher-order network motifs as small connected patterns of higher-order interactions that appear in an observed hypergraph at a frequency that is significantly higher than a suitably randomized system. 

Similarly to what happens with traditional motifs, the steps required to perform a higher-order motif analysis are (i) counting the frequency of each higher-order motif in a network, (ii) comparing the frequency of each motif with that observed in a null model, and (iii) evaluating their over- or under-expression using a statistical measure. Counting algorithms for traditional motifs mistreat by design information about groups interactions, since they are not able to capture patterns of hyperlinks. A detailed description of our proposal for algorithms and tools able to extract and evaluate higher-order motifs is reported in the Methods section.

For our motif analysis of real-world higher-order systems, we collected a number of freely available networked datasets. The datasets~\cite{justice2019database, pacs_data, mastrandea2015contact, benson2018simplicial, chaintreau2007impact, kunegis2013konect, genois2018can, gelardi2020measuring, sinha2015MAG, leskovec2010signed, leskovec2010predicting, pinero2019disgenet, pinero2016information, pinero2015disgenet, queralt2015disgenet-rdf, bauer2011gene-disease, genois2015data, vanhems2013estimating} come from a variety of domains: social (proximity contacts, votes), technology (e-mails), biological (gene/disease, drugs) and co-authorship (See Supplementary Information). In some datasets, higher-order structures are naturally encoded as hyperlinks (e.g. three authors collaborating on the same paper), in others we infer higher-order structures from pairwise interactions (e.g. for face-to-face interactions recorded over time, we promote cliques of size $k$ to hyperlinks of order $k$ if the corresponding three dyadic encounters happened at the same time).

\subsection{Combinatorial analysis of higher-order motifs}

\begin{figure*}[htp]
    \centering
    \includegraphics[width=\linewidth]{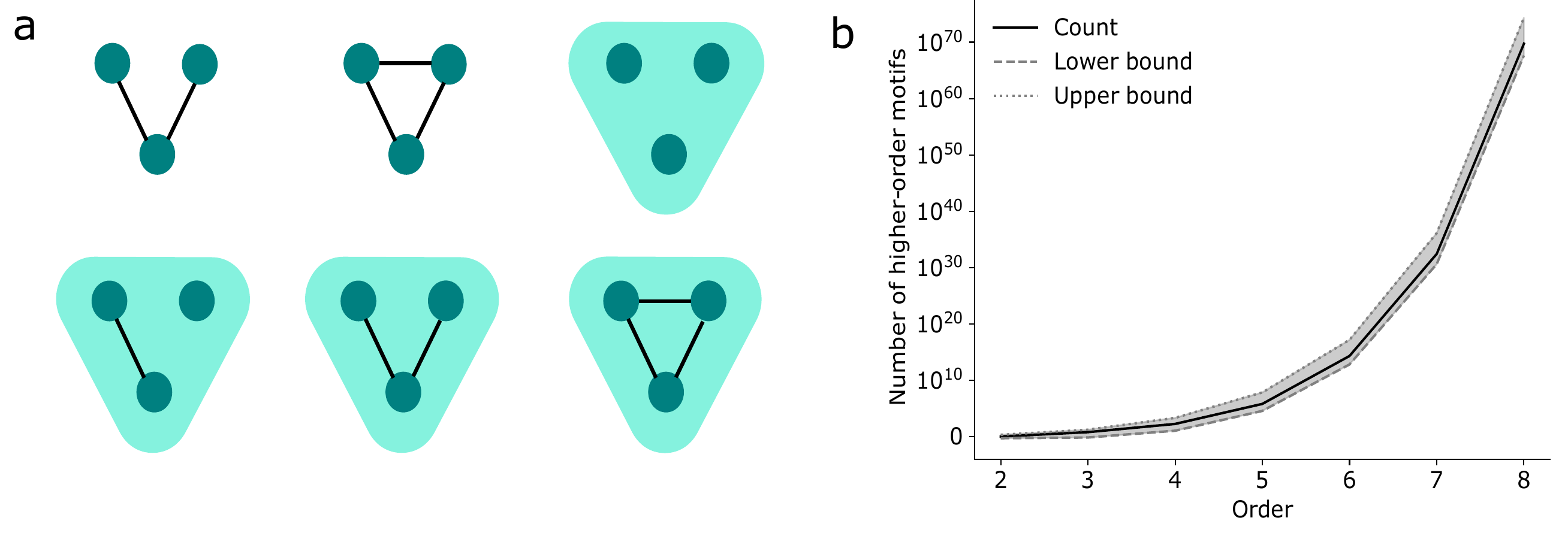}
    \caption{\textbf{Combinatorics of higher-order motifs.} \textbf{a)} Enumeration of all the six possible patterns of higher-order interactions involving three nodes. \textbf{b)} Analytical upper and lower bounds on the number of higher-order motifs as a function of the order.}
    \label{fig:1a}
\end{figure*}

The number of possible patterns of pairwise undirected interactions involving three connected nodes is only two, however it grows to six when considering also higher-order interactions (Fig.~1a). Finding an analytical form encoding the dependence of the number of higher-order motifs on the motif order $k$ is a challenging task due to the constraints related to the computation of all possible combinations of higher-order interactions among $k$ nodes. However, we are able to compute a lower- and upper-bound for this number. We denote with $m$ the number of all the possible non-isomorphic connected hypergraphs of $k$ vertices (we recall that two hypergraphs are isomorphic if they are identical modulo relabeling of the vertices). To compute an upper bound on $m$, we can count the number of labelled hypergraphs ignoring the constraint on being non-isomorphic and connected. There are $\binom{k}{i}$ possible hyperedges of size $i$ over $k$ vertices. We are interested only in the hyperedges with cardinality at least $2$, therefore there are $\sum_{i=2}^{k} \binom{k}{i} = 2^k - k - 1$ possible hyperedges. When creating a labelled hypergraph we can either include each hyperedge or not, this yields a total number of possible labelled hypergraphs equal to $2^{2^k - k - 1}$. To compute the lower bound of $m$, we construct connected hypergraphs on $k$ vertices as follows. First, we pick any chain of edges and put all the edges in the hypergraph. This uses $k-1$ edges and makes sure the hypergraph is connected. There are $(2^k - k - 1) - (k - 1) = 2^k - 2k$ potential edges left over. For each of those edges, we can add them or not to the hypergraph, yielding at least $2^{2^k - 2k}$ connected hypergraphs. However, we have to count only non-isomorphic copies, and have so far counted labelled graphs. For each unlabelled graph, there are at most $k!$ ways of labeling the vertices. So the number of non-isomorphic connected hypergraph is at least $\frac{2^{2^k - 2k}}{k!}$. In Fig.~1b we report the upper- and lower-bound on the growth of the possible higher-order motifs as a function of the order, as well as the exact count for small orders, showing that this function has a super-exponential growth. The combinatorial explosion of higher-order motifs makes intractable their storing and indexing in memory for high orders, which are necessary steps to count their occurrences in empirical hypergraphs and evaluate their over- or under-expression. Given these combinatorial difficulties, in the following we focus on the analysis of the higher-order motifs of order $3$ and $4$.

\subsection*{Motifs of order $3$}

\begin{figure*}[htp]
    \centering
    \includegraphics[width=\linewidth]{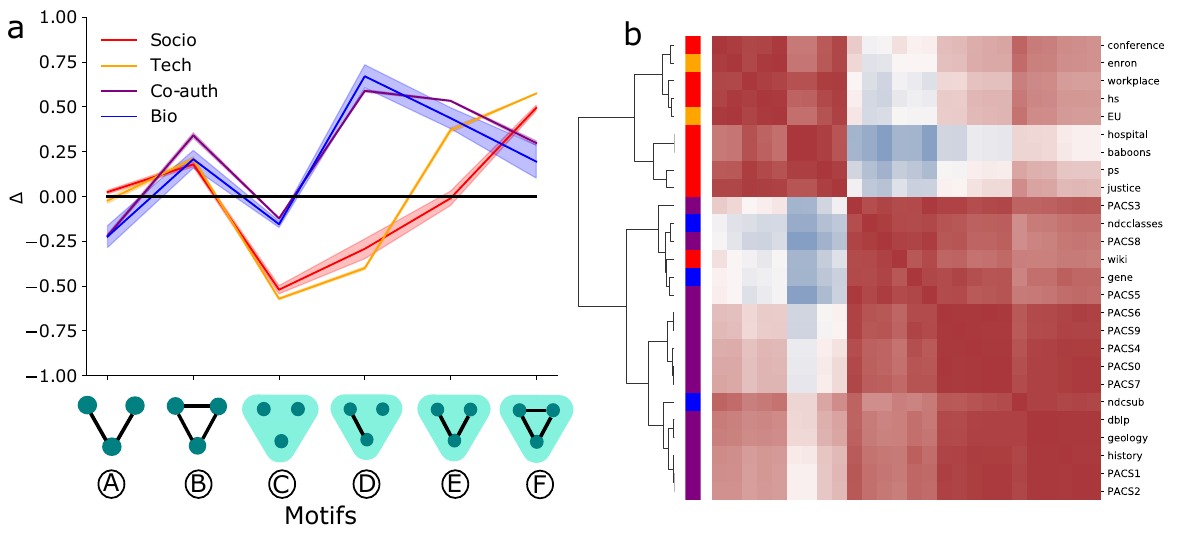}
    \caption{\textbf{A higher-order fingerprint for hypergraphs at the network microscale.} \textbf{a)} Significance Profiles (SP) of hypergraphs from higher-order motifs of order $3$. Over-expressed higher-order motifs are associated to specific functionalities of the system. To simplify the plot, we averaged and grouped higher-order motif profiles of networks from the same domain. For each domain, we represent the mean of the respective higher-order motif profiles with a solid line and the standard error of the mean with a shaded area. \textbf{b)} Correlation matrix of the investigated datasets computed on SPs. SPs of networks from similar domains or that share functionalities display a positive correlation. We identify two large higher-order families of hypergraphs, characterized by distinct higher-order connectivity patterns at the local scale. In both the panels, the domain of every network is labeled with different colors: red for the social domain, orange for e-mails, purple for the co-authorship domain and blue for the biological domain.}
    \label{fig:1b}
\end{figure*}

The over- and under-expression measures of each higher-order motif in a hypergraph are concatenated in a significance profile (SP, see Methods) that constitutes a fingerprint of the local structure of the network. In this section, we characterize the local connectivity of empirical networks at the smallest scale, with higher-order motifs of order $3$.

We compute the SPs of a domain by grouping and averaging the SPs of all networks that belong to it. The analysis of the higher-order profiles of order $3$ highlights the relative structural importance of certain patterns of higher-order interactions (Fig.~2a). The pairwise triangle \Circled{B} appears to be a strong motif in all the clusters, whereas the greatest differences across domains emerge from the motifs involving a $3$-hyperlink and at least one dyadic edge. In the social and technological domains, the motif \Circled{F} made by a 3-hyperlink and a triangle of dyadic edges is strong, suggesting that entities interacting in groups also tend to interact individually. In co-authorship networks, the strongest motifs are \Circled{D} and \Circled{E}, which involve a $3$-hyperlink and one or two dyadic edges, indicating that in these domains there might be a hierarchical structure that prevents all nodes from interacting equally in pairs, as in the case of a research leader that co-authors papers with students and postdocs while the latter do not co-authors papers without the former. A similar motif is also found to be overabundant in biological system. Moreover, SPs allow also to analyze \emph{anti-motifs}, i.e. motifs that are strongly underrepresented. An anti-motif in the social and technological domains is \Circled{C}, the $3$-hyperlink without any dyadic interaction, indicating that it is unlikely that an interaction in group is not followed or preceded by any pairwise interaction. The biological and co-authorship domain do not display any strong anti-motif.

More insights can be obtained by analyzing correlations among significance profiles of different systems. We perform a cluster analysis considering the pairwise correlation between the SP of each dataset as (the opposite of) a distance (Fig.~2b). The analysis shows the emergence of two main clusters, i.e. families of higher-order networks that share similar patterns of higher-order interactions at the microscale. The inferred clusters reproduces the partitions of domains observed in Fig.~2a
(social and technical datasets in a cluster, biological and co-authorship ones in the other), but it also offers a more nuanced view on the similarity across any couple of datasets. %The upper-left cluster incorporates datasets from the social and technology domains. The lower-right cluster incorporates datasets from the biological and co-authorship domains.

\subsection*{Motifs of order $4$}
\begin{figure*}[htp]
    \includegraphics[width=\linewidth]{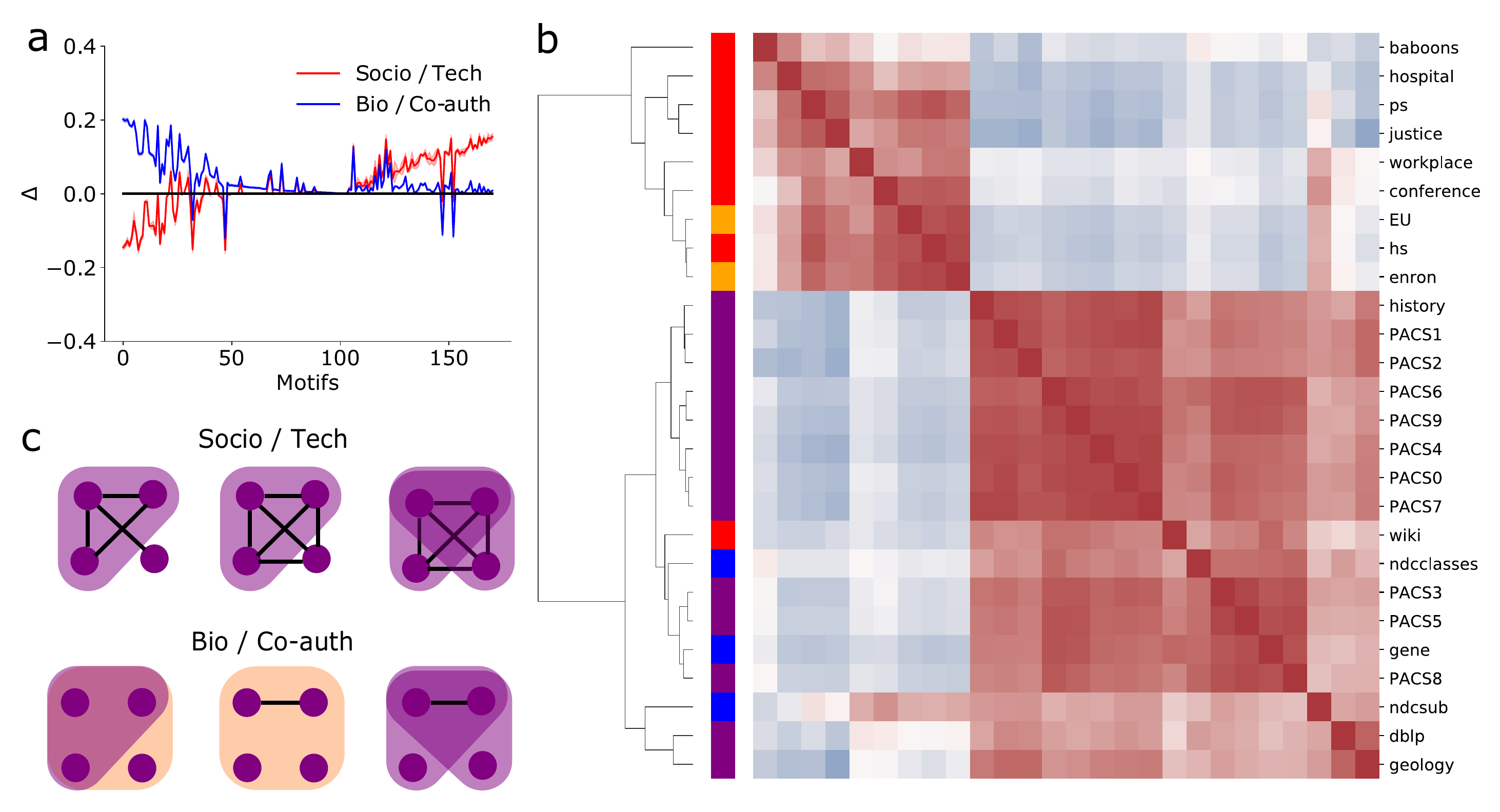}
    \caption{\textbf{Analyzing the local structure of hypergraphs via higher-order motifs of order $4$.} \textbf{a)} Significance Profiles (SP) of hypergraphs from higher-order motifs of order $4$. SPs are much more complex due to the increase in the number of considered patterns of higher-order interactions. We group and average the SPs of networks from the same higher-order family (i.e. Socio/Technological and Co-authorship / Biological) and sort the motifs on the $x$-axis based on their ability to discriminate the two higher-order families. Distinct characteristic higher-order motifs of order $4$ are associated to the two classes of networks. The shaded area represents the standard error of the mean. \textbf{b)} Correlation matrix of the investigated datasets computed on SPs of order $4$. The matrix provides richer information than its equivalent at order $3$ on the local structure of networks: the two big clusters emerge again but are better separated, and display a richer intra-cluster hierarchical structure. \textbf{c)} The six most representative higher-order motifs from the two clusters.}  
    \label{fig:fig2}
\end{figure*}

In the previous section we have systematically investigated the smallest higher-order motifs. 
The number of possible patterns of higher-order interactions involving $4$ nodes is significantly higher than the corresponding with $3$ nodes, as it grows from $6$ to $171$. Despite the difficulties associated to this increase, analysing motifs of order $4$ provides more nuanced information about the local structure of networks with respect to $3$-motifs. 

In Fig.~3a, we group similar clusters together showing the average of their SPs for motifs of order $4$. The motifs on the $x$-axis are sorted in such a way to maximize the distance between the clusters. On the left-end of the $x$-axis, we find motifs that are strongly over-represented in the Biological / Co-authorship domain, while they are under-represented in the Sociological / Technological domain. Conversely, on the right-end of the $x$-axis we find motifs that are over-represented in the Sociological / Technological domain, while not characteristic for the other domain. This observation suggests that both the extremes of the $x$-axis carry information about the structural differences among the clusters.

The richer structural information carried out by the motifs of order $4$ compared to their counterparts of order $3$ is highlighted in the clustering analysis (Fig.~3b). When focusing on the two main clusters, the results are comparable with the previous cluster analysis. However, a richer hierarchical intra-cluster organization naturally emerges, as well as a better separation between the two clusters.

Finally, we characterize the two Sociological / Technological and the Biological / Co-authorship clusters by means of their most over-expressed, and therefore most representative, higher-order motifs of order $4$ (Fig.~3c). The Sociological / Technological domain shows an over-representation of structures involving more lower-order inner relations (e.g. dyadic links), while the Biological / Co-authorship domain displays a preference towards less relations but of higher-order. 

\subsection*{Higher-order motifs and reinforcement}

In order to understand if and how the occurrence of dyadic connections affects the strength of group interactions, we investigate how much the weight of each hyperlink (i.e. the number of times it appears) is correlated with the number of underlying pairwise links. We find that a clear positive trend naturally emerges, indicating the existence of correlation between a rich inner pairwise structure and the weight of a hyperlink (Fig.~4a). We dubbed this phenomenon as higher-order \emph{structural reinforcement}.

%In this section we aim at characterizing how the inner structure of dyadic connections affects the strength of the group interactions in empirical higher-order networks. 

%A first question we can answer is how the frequency of appearance of each group interaction is correlated with the number of underlying pairwise relations. We show that there is a positive correlation between a rich inner pairwise structure and the frequency of a group interaction (Fig.~3a). We dubbed this phenomenon as \emph{structural reinforcement}.

Moreover, we used the metadata about personal relationships between students recorded in the High School dataset from SocioPatterns to understand if a similar reinforcing behavior is observed in the presence of friendships connections between individuals. Friendship data has been collected in two ways, from Facebook accounts and through a questionnaire. In the first case, two students are always reciprocally friends, while in the second case a friendship may be not reciprocal. In Fig.~4b we analyze the relationship between the average number of friends (both on Facebook and by questionnaire) and the topology of the different motifs in the proximity hypergraph. Our findings suggest that the higher is the number of pairwise interactions between students that interact in hyperlinks of size three, the higher will be the number of friends in the group, further corroborating the existence of reinforcement mechanisms.

%Beside data about proximity contacts among high school students, the High School dataset from SocioPatterns offers data about friendships among the students.

\begin{figure*}[htp]
    \includegraphics[width=\linewidth]{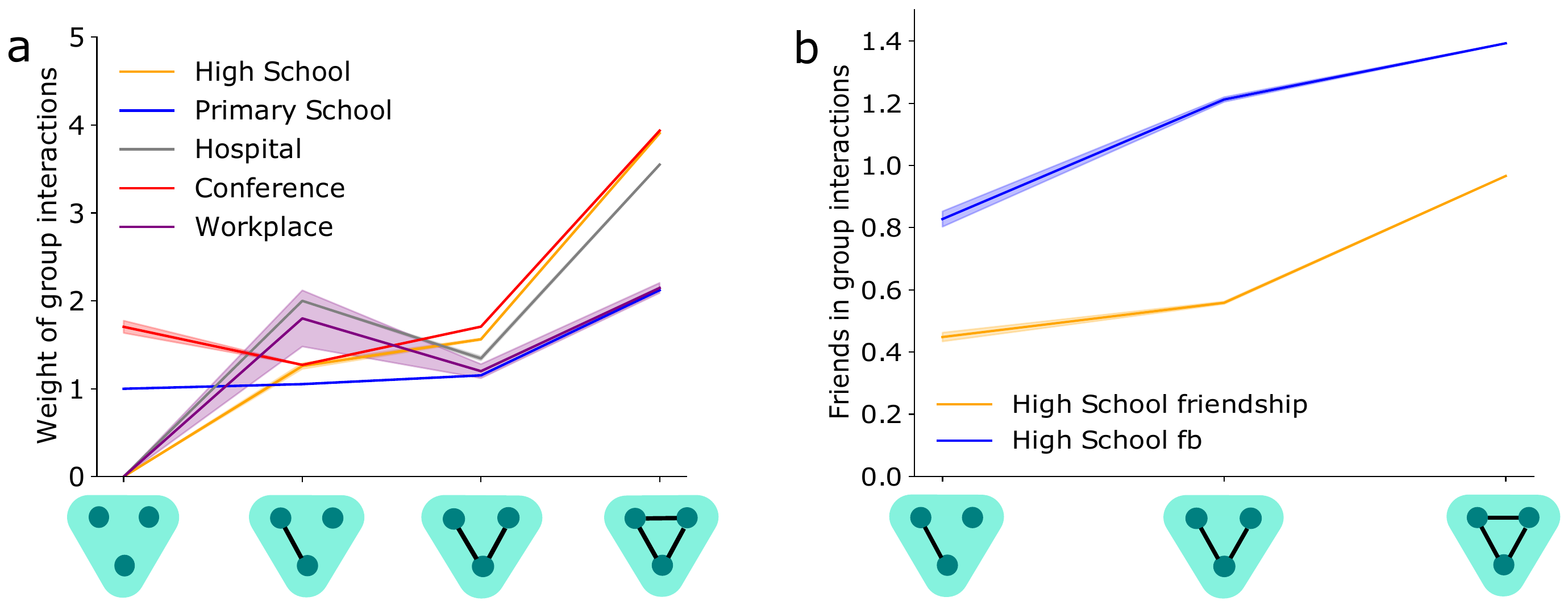}

    \label{fig:fig3}
    \caption{\textbf{Structural reinforcement.} A rich supporting inner structure of pairwise links makes group interactions stronger. In both panels, the stronger levels of connectivity are observed when the number of dyadic connections increases. \textbf{a)} Mean weight of each group interaction (i.e. the number of times it appears) as a function of the number of its nested pairwise links. \textbf{b)} Mean number of friends (certified by a Facebook friendship or by a questionnaire) in group interactions as a function of the number of their nested pairwise links. In both panels, the shaded area represents the standard error of the mean.}

\end{figure*}

\subsection*{Nested organization of higher-order interactions}
A limitation of our work is that an exhaustive approach like the one followed in the previous sections is only feasible for motifs of order $3$ and $4$. As motif analysis becomes highly computationally demanding when the order increases, in the following we focus on characterizing the nested structures of large hyperlinks. This means that, instead of counting the exact frequency of each pattern of higher-order interactions involving a number of nodes higher than $3$ or $4$, we settle for extracting statistics on the hierarchical structures of interactions inside hyperlinks of any size. The advantage of this approach is that it still provides information about the local structure of sub-modules of a network, while its computational complexity is only linear in the number of hyperedges in the hypergraph.

First, we look at the average number of edges in the inner structures of hyperedges of different sizes (Fig.~5a). The networks are grouped according to the clusters detected in the previous analysis. While the cluster Bio / Co-auth does not display evident differences in the number of nested edges with the growth of the size of the hyperedges, the Socio / Tech domain shows a clear growing trend with a change of slope after orders 5 and 6. 

In order to complement this information, we looked at how the mean size of the nested edges changes with the growth of the size of the analyzed hyperedges (Fig.~5b). In this case, both the clusters show a growing trend, with the Bio / Co-auth domain displaying a faster growth. Thus, while Socio / Tech networks tend to have more edges in the inner structure of their large hyperedges, they tend to be of small size. The Bio / Co-auth domain, instead, shows an opposite behavior. All in all, this suggests that, in agreement with previous findings, also at higher scales Socio / Tech network motifs are systematically more nested.

\begin{figure*}[htp]
    \includegraphics[width=\linewidth]{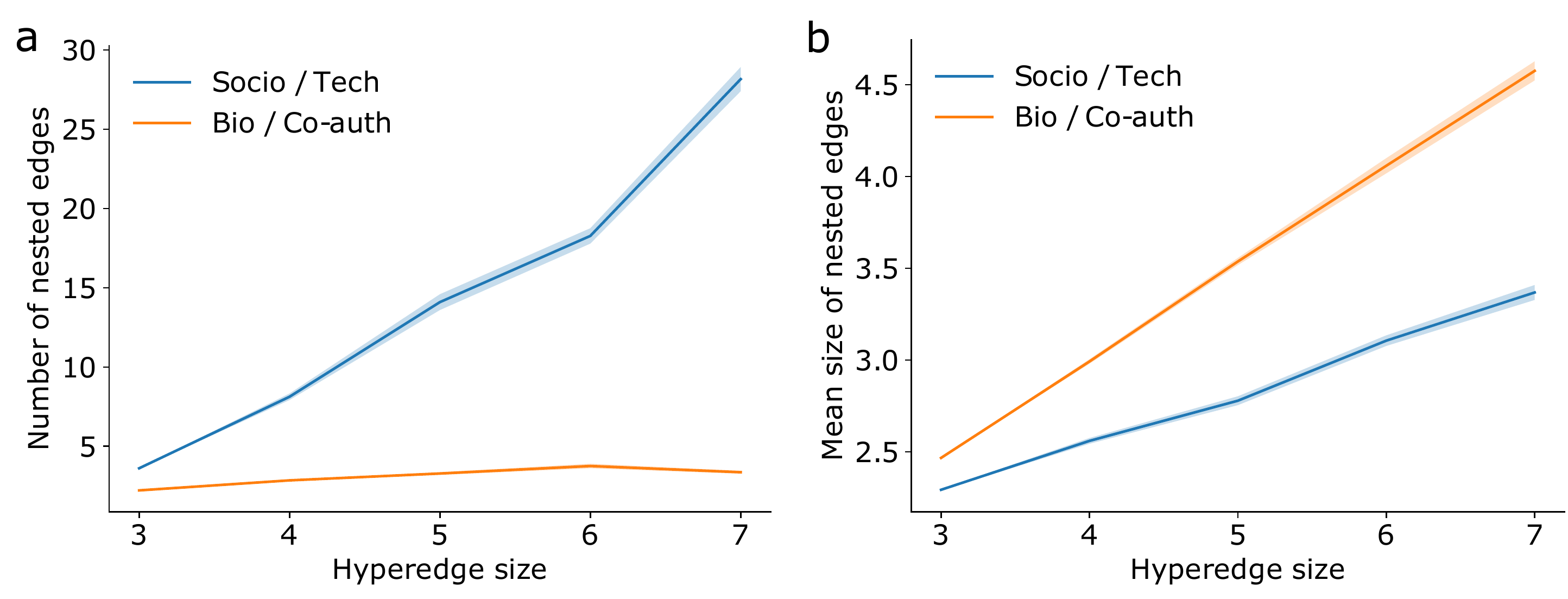}

    \label{fig:fig4}

    \caption{\textbf{Nested organization of group interactions.} Different higher-order families of hypergraphs can display very different hierarchical organization of their higher-order interactions. In both panels, we grouped datasets according to the clusters detected in the previous analysis. \textbf{a)} Mean number of links in the inner structure of the hyperedges as a function of their size. The cluster Bio / Co-auth displays a static behavior, while the Socio / Tech domain shows a clear increasing richness of the inner hierarchical structures of the hyperedges. \textbf{b)} Mean average size of the links in the inner structure of the hyperedges as a function of their size. Both the clusters show a linear growing trend, however the cluster Bio / Co-auth grows faster. All in all, Socio / Tech networks tend to have a lot of small-size edges in the inner structure of their hyperedges. The Bio / Co-auth domain, instead, tend to prefer few large-size edges. In both panels, the shaded area represents the standard error of the mean.}

\end{figure*}

\section*{Discussion}
The framework of network motifs is widely recognized as a fundamental tool for the analysis of complex networks. Able to highlight local structural characteristics of networks and influence their dynamics, motifs can be considered the fundamental building blocks of networks, and have produced applications in a number of fields such as biology and social network analysis.

Modeling complex systems by means of hypergraphs has recently emerged as a fundamental tool in Network Science, prompting the question of how to identify and assess network motifs in the presence of higher-order interactions. With the aim to extract the local fingerprint of hypergraphs, in this work we introduced the notion of higher-order network motifs, which are small, possibly overlapping patterns of higher-order interactions that are statistically over-represented with respect to a null model. We proposed a combinatorial characterization of higher-order network motifs, as well as an efficient algorithm to evaluate their statistical significance on empirical data. These tools allowed us to extract fingerprints of a variety of real-world systems by focusing on their characteristic patterns of higher-order interactions among small groups of nodes, showing the emergence of families of hypergraphs characterized by similar local structures. Moreover, we captured a structural reinforcement mechanism that associates stronger weights of higher-order interactions to groups of nodes that interact more at the pairwise level.

Similarly to the case of traditional pairwise network motifs, we believe that higher-order network motifs can pave the way to applications in a number of domains, pushed by the growing awareness of the relevance of the higher-order nature of interactions in many real-world systems. Given the possible applications of this framework in data-intensive domains, a limitation of our proposed approach is its scalability. In this work, indeed we proposed an algorithm that allows to perform an exhaustive search, and for this reason focuses on higher-order network motifs of size $3$ and $4$. However, we believe there is room for different approaches, which sacrifices exhaustiveness but could allow to gain deep insights on motifs of greater size. As a first step in this direction, we looked at the nested structure of patterns of hyperlinks of larger orders. In addition to this, we believe that the development of sampling methods for the statistical evaluation of higher-order network motifs will be critical for more widespread real-world applications. All in all, our work highlights the informative power of higher-order motifs, providing a first way to extract higher-order fingerprints in hypergraphs at the network microscale      

\section*{Methods}
%\hl{Here the technical details of the algorithms, the measures of significance, etc.}

A higher-order motif analysis involves two steps: counting the frequency of each target higher-order motif in an observed network and comparing them with those of a null model to establish the over- or under-expression of certain subgraph patterns.

To count the frequency of each higher-order motif of order $k$ in a hypergraph, we propose an exact algorithm that combines different steps. A fundamental sub-task to solve efficiently is the hypergraph isomorphism problem (i.e. establishing the equivalence under relabeling of two hypergraphs). In fact, for each occurrence of a connected subgraph with $k$ nodes we need to update the frequency of the respective higher-order motif of order $k$. This problem can be solved efficiently by enumerating and indexing all the higher-order motifs of order $k$ with all the respective relabelings, allowing to update and count occurrences of connected subgraphs in constant time. Since we are interested only in patterns of small subgraphs of order $3$ and $4$ this is doable: the number of possible non-isomorphic patterns of higher-order interactions involving $4$ nodes is $171$, a number that makes them storable in memory. To enumerate subgraphs we use an algorithm that proceeds in a hierarchical way. It first iterates over all the hyperlinks of size $k$, which are able to directly induce a motif. Then it iteratively considers hyperlinks of lower orders until it reaches the traditional dyadic links. Since hyperlinks of order lower than $k$ are not able to directly induce a motif, the algorithm proceeds in a way similar to~\cite{wernicke2006esu} and selects the remaining nodes by considering the neighborhood of the subgraph. Once selected $k$ nodes, to efficiently construct their induced subgraph we iterate over the power set of the $k$ nodes (which corresponds to $2^n$ possible hyperedges) and keep only the hyperedges that exist in the original hypergraph.

As a null model, we use the configuration model proposed in~\cite{chodrow2020configuration}. We sample from the configuration model $n = 20$ times and compute the frequencies of the higher-order motifs in each sample. To validate the over- and under-expression of certain patterns, we use the abundance of each motif $i$ relative to random networks proposed in~\cite{milo2004superfamilies}.

\begin{equation}
    \Delta_i = \frac{N\text{real}_i - \langle N\text{rand}_i \rangle}{N\text{real}_i + \langle N\text{rand}_i \rangle + \epsilon}
\end{equation}

Without loss of generality we set $\epsilon = 4$, again following~\cite{milo2004superfamilies}. From the same reference, we define the Significance Profile (SP) as the vector of $\Delta_i$ of a network normalized to length 1.

\begin{equation}
    \text{SP}_i = \frac{\Delta_i}{\sqrt{\sum \Delta_i^2}}
\end{equation}

The code for higher-order motif analysis will be freely available after publication.

\bibliography{biblio}

\end{document}

\title{Supplementary Information\\ {\large Higher-order motif analysis in hypergraphs}}

\author{Quintino Francesco Lotito}
\author{Federico Musciotto}
\author{Alberto Montresor}
\author{Federico Battiston}

%\flushbottom
\maketitle

\section{Datasets}
In this section we present additional information on the networked datasets gathered to perform our motif analysis of real-world higher-order systems. The datasets come from a variety of domains: social (proximity contacts, votes), technology (e-mails), biological (gene/disease, drugs) and co-authorship.

\begin{itemize}
    \item \textbf{Gene/disease.} Nodes correspond to genes and each hyperedge is the set of genes associated to a disease.
    \item \textbf{NDC\_classes.} Nodes correspond to classification labels and each hyperedge is the the of all the labels applied to a drug. Each drug is timestamped with the day it was first marketed.
    \item \textbf{NDC\_substances.} Nodes correspond to substances and each hyperedge is the set of substances in a drug. Each drug is timestamped with the day it was first marketed.
    \item \textbf{DBLP.} Nodes correspond to authors and each hyperedge is the set of authors on a scientific publication tracked by DBLP. Each paper is timestamped with the year of publication.
    \item \textbf{History.} Nodes correspond to authors and each hyperedge is the set of authors on a scientific publication in the field of history. Each paper is timestamped with the year of publication.
    \item \textbf{Geology.} Nodes correspond to authors and each hyperedge is the set of authors on a scientific publication in the field of geology. Each paper is timestamped with the year of publication.
    \item \textbf{PACS.} Nodes correspond to authors and each hyperedge is the set of authors on a scientific publication in the field of physics. Each paper is timestamped with the year of publication and with its PACS classification. The latter allows to split the set of papers in ten subfields of physics by using the highest hierarchical level of PACS classification.  
    \item \textbf{Wiki.} This dataset contains all the Wikipedia voting data from the inception of Wikipedia until January 2008. Nodes correspond to Wikipedia users and each hyperedge is the set of users that expressed the same vote in a voting event.
    \item \textbf{email-EU.} Nodes correspond to users in a European research institution and each hyperedge consists of the sender and all recipients of an email. Each email is timestamped.
    \item \textbf{email-ENRON.} Nodes correspond to Enron employees and each hyperedge consists of the sender and all recipients of an email. Each email is timestamped.
    \item \textbf{Justice.} This dataset records all the votes expressed by the justices of the Supreme Court in the US from 1946 to 2019 case by case. Nodes correspond to justices and each hyperedge is the set of justices that expressed the same vote in a case.
    \item \textbf{Primary school.} Nodes are students of a primary school. Wearable sensors are exploited to construct a network of proximity contacts among the students. Contacts are aggregated in time-windows of $20$ seconds. Each hyperedge is a maximal clique in each layer (i.e. each interval) of the temporal network of contacts. 
    \item \textbf{High school.} Nodes are students of a high school. Wearable sensors are exploited to construct a network of proximity contacts among the students. Contacts are aggregated in time-windows of $20$ seconds. Each hyperedge is a maximal clique in each layer (i.e. each interval) of the temporal network of contacts. 
    \item \textbf{Conference.} Nodes are participants of a conference. Wearable sensors are exploited to construct a network of proximity contacts among people. Contacts are aggregated in time-windows of $20$ seconds. Each hyperedge is a maximal clique in each layer (i.e. each interval) of the temporal network of contacts.
    \item \textbf{Hospital.} Nodes are people at a hospital. Wearable sensors are exploited to construct a network of proximity contacts among people. Contacts are aggregated in time-windows of $20$ seconds. Each hyperedge is a maximal clique in each layer (i.e. each interval) of the temporal network of contacts.
    \item \textbf{Workplace.} Nodes are employees of a company. Wearable sensors are exploited to construct a network of proximity contacts among people. Contacts are aggregated in time-windows of $20$ seconds. Each hyperedge is a maximal clique in each layer (i.e. each interval) of the temporal network of contacts.
    \item \textbf{Baboons.} Nodes are baboons in an enclosure of a Primate Center in France. Wearable sensors are exploited to construct a network of proximity contacts among baboons. Contacts are aggregated in time-windows of $20$ seconds. Each hyperedge is a maximal clique in each layer (i.e. each interval) of the temporal network of contacts.
    
\end{itemize}

The summary statistics of the datasets are reported in Table S1.

\begin{table*}
\centering
\begin{tabular}{l|r|r|r|r|r|r|r|l}
\toprule
Dataset & N & E & $E_2$ & $E_3$ & $E_4$ & $E_5$\\
\midrule
Gene/disease & 9703 & 11181 & 1311 & 614 & 443 & 363 \\
NDC\_classes & 1161 & 1088 & 297 & 121 & 125 & 94 \\
NDC\_substances & 5311 & 9906 & 1130 & 745 & 535 & 500\\
DBLP & 1924991 & 2466799 & 693363 & 667291 & 419434 & 205970\\
History & 1014734 & 895439 & 160885 & 47423 & 19120 & 8775\\
Geology & 1256385 & 1203895 & 275736 & 227950 & 159509 & 99140\\
PACS0 & 98478 & 75985 & 12168 & 8373 & 14068 & 3064 \\
PACS1 & 67055 & 43957 & 10532 & 7869 & 4647 & 2015 \\
PACS2 & 47475 & 27504 & 5424 & 3452 & 2224 & 1411 \\
PACS3 & 33479 & 16977 & 2099 & 2105 & 2590 & 1169 \\
PACS4 & 57533 & 41133 & 8977 & 6571 & 5139 & 2897 \\
PACS5 & 14455 & 5306 & 1062 & 1041 & 719 & 477 \\
PACS6 & 71989 & 41663 & 5140 & 3293 & 4946 & 1715 \\
PACS7 & 82964 & 55227 & 9664 & 9325 & 8294 & 4915 \\
PACS8 & 13451 & 5618 & 1659 & 1447 & 910 & 485 \\
PACS9 & 18666 & 7515 & 2361 & 2143 & 1034 & 397 \\
Wiki & 6210 & 5925 & 593 & 427 & 338 & 304 \\
email-EU & 998 & 25027 & 12753 & 4938 & 2294 & 1359 \\
email-ENRON & 143 & 1512 & 809 & 317 & 138 & 63 \\
Justice & 38 & 2864 & 216 & 456 & 506 & 560 \\
Primary school & 242 & 12704 & 7748 & 4600 & 347 & 9 \\
High school & 327 & 7818 & 5498 & 2091 & 222 & 7 \\
Conference & 403 & 10541 & 8268 & 1861 & 258 & 63 \\
Hospital & 75 & 1825 & 1108 & 657 & 58 & 2 \\
Workplace & 92 & 788 & 742 & 44 & 2 & 0 \\
Baboons & 13 & 231 & 78 & 142 & 11 & 0 \\
\bottomrule
\end{tabular}
\caption{Details of the real-world networked datasets considered for our experiments. Each higher-order network is described by its number of nodes, its total number of hyperedges and its number of hyperedges of size $2$, $3$, $4$ and $5$.}
\label{tab:datasets}
\end{table*}